# Whole body dynamic PET kernel reconstruction using nonnegative matrix factorization features

Alan Miranda[1,2] and Steven Staelens[1,2]

[1] Molecular Imaging Center Antwerp, University of Antwerp, Antwerp, Belgium

[2] µNeuro Research Centre of Excellence, University of Antwerp, Antwerp, Belgium

**Abstract**

The kernel reconstruction is a method that reduces noise in dynamic positron emission tomography (PET) by exploiting spatial correlations in the PET image. Although this method works well for large anatomical regions with relatively slow kinetics, whole body PET reconstruction with the kernel method can produce suboptimal results in regions with fast kinetics and high contrast. In this work we propose a new design of the spatial kernel matrix to improve reconstruction in fast and slow kinetics body regions. We calculate voxels features using nonnegative matrix factorization (NMF) with optimal rank selection. These features are then used to calculate similarities between voxels considering relative differences between features to adapt to a wide range of activity levels. Simulations and whole body mouse scans of high temporal resolution [$^{18}$F]SynVesT-1, low dose [$^{11}$C]raclopride, and [$^{18}$F]Fallypride were performed to assess the performance of the method in different settings. In simulations, bias vs variance tradeoff and contrast was improved using the NMF kernel matrix, compared with the original kernel method. In real data, fast kinetic regions such as the heart, veins and kidneys presented oversmoothing or artifacts with the original kernel method. Our proposed method did not present these effects, while reducing noise. Brain kinetic modeling parametric maps with image derived input function ([$^{18}$F]SynVesT-1) and with reference region ([$^{11}$C]raclopride and [$^{18}$F]Fallypride) also had lower standard error using the proposed kernel matrix compared with other methods. The NMF kernel reconstruction reduces noise and maintains high contrast in whole body PET imaging, outperforming the traditional kernel method.

## 1. Introduction

In order to perform kinetic modeling quantification of positron emission tomography (PET) radiotracers, the time course of radiotracer uptake has to be followed from the time injection to (typically) a couple of hours after administration. Dynamic PET reconstruction then has to be calculated with this data. Since usually short time frames need to be reconstructed to capture the fast kinetics of the radiotracer uptake, dynamic PET reconstructions can suffer from high noise. To reduce noise in dynamic PET reconstruction, a wide variety of methods have been developed, relying for example in spatial and/or temporal regularization, and using prior information, either from the expected tracer kinetics (e.g. direct reconstruction), or using other imaging modalities (e.g. MRI and CT) [1-3]. More recently machine learning methods have also been explored for noise reduction in PET image reconstruction [4, 5].

The kernel method is one of the PET reconstruction techniques used to reduce noise in dynamic PET [3]. In this method, by finding spatial correlations between voxels, spatial basis functions can be constructed to represent the PET image with lower noise. The popularity of the kernel method stems from its relative ease of calculation, outperforming more complex algorithms [3, 6], allowing its implementation using a kernelized version of the also popular maximum likelihood expectation maximization (ML-EM) algorithm. Several improvements to the kernel method have been developed in recent years. Using anatomical imaging with typically lower noise than PET (e.g. MRI), the (spatial) kernel matrix can be further improved using the anatomical image as prior information [7, 8]. Since introduction of multimodality images can introduce bias in structures present only in the functional PET data, methods to correct for this effect have also been developed [9-11]. In addition to the spatial kernel matrix, including a temporal kernel matrix, exploiting correlations between time frames further reduces noise in dynamic PET images [8, 12].

Machine learning algorithms have also been used together with the kernel method for PET image reconstruction. To avoid the need of additional anatomical scan acquisitions, foundational anatomical medical imaging models have been used to calculate the spatial kernel matrix [13]. The convolutional neural network U-Net has been used to improve the feature extraction from high quality PET images to then calculate the kernel matrix [6], or to regularize the kernel matrix coefficient vector calculated in the kernelized ML-EM [14].

In this work we focus on improvement of the functional kernel matrix, i.e. the spatial kernel matrix constructed from individual subject PET data only, for dynamic PET reconstruction. In terms of practicality, this is the most optimal kernel method since no other modality scans are needed, and no multi subject data is necessary, also reducing the risk of over-fitting bias from other subjects.

For calculation of the kernel matrix in PET dynamic reconstruction, the standard approach is to calculate high quality prior images by reconstructing a few long time frames (composite frames of several minutes), using then these images as voxel features to find voxel correlations [3]. In our experience this approach is suboptimal when structures with highly different kinetics are present in the dynamic PET data, such as in whole body PET. For example, in some radiotracers the heart and kidneys peak uptake last a few seconds or minutes followed by a fast washout, while in other organs such as brain and liver a slow washout could be observed. Using composite frames with duration of several minutes, although reflecting the kinetics of slow washout organs, would average out the kinetics of fast uptake organs.

Another situation where the use of composite frames is suboptimal is when voxels in an organ share the same kinetic profile, with differences only in the amplitude of the uptake. In this case only a single frame for the whole duration of the scan would be optimal to search for correlations between voxels, i.e. searching for similarities only based on voxel intensity magnitude. In comparison, using multiple composite frames with higher noise (compared with a single frame) would introduce additional noise in the calculation of voxels similarities.

Finally, another issue that can arise in the kernel method is oversmoothing of small high contrast regions. This can happen when the kernel matrix coefficients (spatial basis functions) of the high contrast region include voxels with highly different uptake magnitude. A simple approach to correct for this is to set a threshold on the weights calculated with the radial basis function kernel (RBFK) to remove small weights corresponding to lower voxels feature similarity [3]. However, setting too many weights to zero can compromise the smoothing/regularization property of the kernel matrix.

To address these issues here we propose several modifications to the kernel method detailed below. We validated our approach in 2D phantom simulations, and real whole body mouse scans. To demonstrate the applicability of our approach in different scenarios, we applied our method on 3 different preclinical mouse whole body scans: (i) A [$^{18}$F]SynVesT-1 high temporal resolution reconstruction, (ii) a low dose/high noise [$^{11}$C]raclopride scan, and (iii) a [$^{18}$F]Fallypride scan where high contrast in the reference region is necessary.

## 2. Methods

2.1 Nonnegative matrix factorization in PET imaging

Nonnegative matrix factorization (NMF) is a popular unsupervised learning algorithm used in a wide range of fields [15]. NMF seeks to solve the following problem: given matrix $X \in \mathbb{R}^{M \times N}$, find matrices $W \in \mathbb{R}_+^{M \times r}$ and $H \in \mathbb{R}_+^{r \times N}$, where $r$ is the rank of the factorization, such that $X \cong WH$. Usually, $r \ll M, N$, and with proper selection of $r$ (see below, rank selection) noise can be factored out from $X$. Compared to other matrix factorization methods, such as singular value decomposition (SVD), and independent component analysis (ICA), NMF is remarkable in producing results with strong interpretability [15].

In the context of dynamic PET, if matrix $X$ is created by arranging in columns the time activity curves (TACs) of all voxels, columns of matrix $W$ represent the unique TACs that describe all "pure" kinetic profiles, while the rows of $H$ are the weight factors of every voxel for the respective unique TAC. If the rows of $H$ are rearranged in an image, one can reduce the dynamic PET image to its smallest representation (dimensionality reduction) without losing information about the radiotracer kinetics.

Several algorithms and improvements have been (and are still) developed to solve the NMF problem [16]. In this work we use the method from Leplat, et al. [17] which minimizes the following objective function:

$$\underset{W,H}{\text{argmin}} \|X - WH\|_F^2 + \lambda \text{logdet}(W^T W + \delta I) \qquad (1)$$

subject to $W \geq 0$, $H \geq 0$, $H^T 1_r \leq 1$, where $T$ denotes matrix transpose, $1_r \in \mathbb{R}^{r \times 1}$, and logdet() is the logarithm of the determinant. The second term in (1) is a regularization term proportional to the volume of the convex hull created by the columns of $W$ and the origin [17, 18]. This regularization has been found to improve identifiability of NMF solutions, allowing to also recover the pure endmembers (columns of $W$) even when they are absent in the data [18]. The constraint $H^T 1_r \leq 1$, i.e. the constraint that the proportions of each pure TAC in every voxel sum at most to 1, helps to find solutions to $W$ with unambiguous scaling. In our tests, the pure TACs, i.e. the columns of $W$, found with this constraint have scaling close to the true activity of the TACs in the data. Parameter $\lambda$ controls the balance between the data fidelity term (first term in 1) and regularization, while $\delta$ lowers the bound of the logdet() function [18]. The solution to (1) is found using the projected fast gradient method [17], with initialization using the successive projection algorithm [19]. We used the default values proposed by the authors with $\lambda = 0.1$, and $\delta = 0.1$ in all experiments. A Matlab implementation of the algorithm is available from the authors [17].

## 2.2 Rank selection for NMF solutions

Selection of $r$ in NMF is a nontrivial problem [20, 21] and several methods used to estimate it offer different performance depending on the application and the noise level. In our application, we found no single rank selection method produced optimal solutions, and therefore we use a hybrid approach where we use several rank selection algorithms and calculate a weighted score based on all solutions. The following rank selection algorithms were used:

(i) Cross-validation by data imputation [20, 22]: cross-validation is widely used in machine learning algorithms for model selection and tuning of parameters and works by keeping one subset of the data for training, while using the rest as test data to evaluate the learning performance on unseen data. Since removing entire data rows or columns can introduce bias in the NMF solution, one alternative to apply cross-validation for NMF is to mask out random elements of matrix $X$ with a random speckle pattern (kept elements are the train set), apply NMF on the masked-in $X$, and reconstruct the masked-out elements (test set) with the solution from the masked-in $X$. The cross-validation error $CV(r)$ is then the norm-2 of the difference between real and reconstructed data evaluated only in the test set matrix elements. We applied this approach by masking 20% of the elements in $X$, running 10 different realizations to obtain an average cross-validation error for every $r$.

(ii) Solution stability across realizations. As $r$ increases over the optimal rank, NMF solutions start to fit noise, and the results over different random realizations start to differ between each other. Using the training set solutions calculated for cross-validation over all realizations, we calculated the correlation between the rows of $H$ for every pair of realizations and calculated the average correlation among all realization's pairs to obtain $RC(r)$ (realizations correlation). Since the ordering of rows in $H$ can differ between solutions, we first ordered rows of $H$ based on correaltion between columns of $W$ before calculating the correlation of rows in $H$. The optimal rank is calculated as the inflection point in the plot of $r$ vs $RC(r)$, i.e. there is large drop in $RC(r)$ when noise starts to be fitted in the NMF solutions. $RC(r)$ for $r = 1$ is set to 1.

(iii) Prediction strength [23]: clustering prediction strength is calculated by dividing data in train and test sets, calculating clusters using the training ($X^{train} \cong W^{train}H^{train}$) and test ($X^{test} \cong W^{test}H^{test}$) sets separately. Differently from cross-validation, $X^{train}$ and $X^{test}$ are obtained by random sampling of the columns in $X$. To obtain hard clustering from NMF, necessary to calculate the prediction strength, we simply calculate the index $c \in [1..r]$ of the maximum value in the columns of $H^{test}$ to obtain $\mathcal{H}^{test}$. Then for each cluster $c$ in the test set, their points are clustered using the training model. This is done by projecting the columns of $X^{test}$ onto $W^{train}$ to obtain $\mathcal{H}^{test\prime}$. Considering all elements of cluster $c$ in $\mathcal{H}^{test}$, the number of pair of indices in $\mathcal{H}^{test}$ that are in the same cluster according to $\mathcal{H}^{test\prime}$ is divided by the total number of indices pairs. The maximum of this ratio among all test clusters $c$ is the prediction strength $PS(r)$ and has a maximum of 1 when all pair of test points are kept in the same cluster using the training model. In this case, we set 20% randomly selected columns in $X$ as test set and the remaining as training set, also taking the average over 10 realizations. Prediction strength with a single cluster ($r = 1$) is set to 1.

(iv) Noise in pure TACs. Similar to the solution stability across realizations, as noise starts to be fitted at increasing $r$'s, noise in the calculated pure TACs (columns of $W$) also starts to increase. We use the total variation of the pure TACs as noise metric, averaging the value obtained at each column, to get $TN(r)$ (TACs noise). Again, the optimal rank is calculated as the inflection point in the plot of $r$ vs $TN(r)$.

To calculate the final optimal rank we change $PS(r)$ into an error by setting $PS'(r) = 1 - PS(r)$. Then both $PS'(r)$ and $CV(r)$ are scaled to the range 0-1 based on their min-max range to obtain $PS''(r)$ and $CV'(r)$. The total rank error is $TE(r) = aPS''(r) + bCV'(r)$, where $a$ and $b$ where empirically set as 0.2 and 1, respectively. To account for $RC(r)$ and $TN(r)$, the index of the inflection points $r^{RC}$ and $r^{TN}$ are calculated and we subtract a constant, empirically set as 0.2 to the total error at $TE(r = r^{RC})$ and $TE(r = r^{TN})$. The optimal $r$ is finally $r^{opt} = \min_r TE(r)$. The combination of all different rank selection metrics offered a robust method that in our tests gave good results for different radiotracers and noise levels. In all experiments we run the rank test from $r = 1 \dots 6$.

## 2.3 PET kernel reconstruction

The PET projection data $y_i$ ($i = 1 \dots I$ detector pairs) can be modeled using the projection matrix $P \in \mathbb{R}^{I \times J}$ applied on the activity image $x_j$ ($j = 1 \dots J$ voxels):

$$y = Px + r \qquad (2)$$

where $r$ are background events per detector pair. In the kernel method for dynamic PET reconstruction, voxel features $f_j$ can be calculated by reconstructing the data in few composite frames of longer duration than the dynamic frames, where the voxel intensity at every composite frame is the feature value in the corresponding dimension. The radial basis function kernel (RBFK) was proposed to calculate the similarity between voxels using the Euclidian distance between features [3]:

$$\kappa_{jl} = \exp\left(-\frac{\|f_j - f_l\|^2}{2\sigma^2}\right) \qquad (3)$$

where $\sigma$ scales the distance between features to assign a significant weight only when the distance is below a certain threshold. Features at every dimension are normalized by the standard deviation of all voxel features, using then $\sigma = 1$. To reduce the size of the kernel matrix, voxels similarities are searched in a close neighborhood and only the closest $k$ neighbors ($k$NN based on Euclidian distance) are considered in the kernel matrix:

$$K_{jl} = \begin{cases} \kappa_{jl}, & f_l \in k\text{NN of } f_j \\ 0, & \text{otherwise} \end{cases} \qquad (4)$$

Finally to improve performance, the rows of the kernel matrix are normalized to sum 1:

$$\overline{K}^{comp} = \text{diag}^{-1}[K1_J]K \qquad (5)$$

where $comp$ refers to the use of composite frames as voxel features. The kernelized ML-EM algorithm can then be used to estimate the coefficient vector $\alpha$:

$$\alpha^{n+1} = \frac{\alpha^n}{K^T P^T 1_I} K^T P^T \frac{y}{PK\alpha^n + r} \qquad (6)$$

and the PET image is finally $x = K\alpha$. In all cases, $\overline{K}^{comp}$ was calculated by dividing the total scan in 6 composite frames of equal length, with 48 $k$NN, and setting a threshold of 0.9 in the RBFK to improve contrast.

## 2.4 Extraction of voxel features using nonnegative matrix factorization

To capture richer kinetic information, we use the ML-EM independent frame dynamic reconstruction (IFDR) with the original framing to calculate voxels features.

*Body segmentation.* As a initial step, we perform hard clustering of the voxels based on tracer kinetics similarities in the IFDR. This is done to separate body regions which are not likely to share kinetics profiles. For example, the bladder and the brain are not expected to have similar radiotracer kinetics. Briefly, IFDR is first filtered in the temporal dimension using SVD, removing small singular values components. Then, dimensionality reduction of IFDR is performed using NMF as described above, followed by non-local means spatial filtering. Voxels of reduced dimensionality then serve as input to the gaussian mixture model algorithm to calculate hard clustering. Matlab functions "fitgmdist" and "cluster" were used for this. Finally, these clusters are refined using a modified version of the SLIC algorithm [24]. This algorithm showed good performance with several radiotracers and different noise levels.

*Voxel features calculation.* Using voxel clusters labels $c = 1, \ldots, C$, data matrix $X^c$ is created concatenating only voxels TACs from cluster $c$. The NMF rank selection algorithm is run for this data matrix, and $W^c$ and $H^c$ are calculated with $r^{opt,c}$ components. In the following, we obtained better results by scaling the rows of matrix $H^c$ by the L1-norm of the corresponding column in $W^c$, obtaining $\bar{H}^c$. In this way, components associated with higher activity pure TACs have larger weight. Columns of matrix $\bar{H}^c$ are the $d$-dimensional voxel features $f_j^d$, where $d = r^{opt,c}$ can be different between voxels clusters.

## 2.5 Kernel matrix calculation with NMF voxels features

Similarities between voxels are searched only in voxels within the same cluster. Then, instead of using the Euclidian distance to calculate distance between features, we consider the voxel features as vectors that represent the shape and magnitude of the kinetic profile. Feature vectors pointing in the same direction have the same kinetic profile shape, while the magnitude of the vector defines the activity scale difference between kinetic profiles. Therefore, to separate similarities based on kinetic profile shape and magnitude, we first normalize the voxels features to the unit $d$-dimensional sphere, i.e. $\hat{f}_j^d = f_j^d / \|f_j^d\|$. Then, we define the RBFK for shape and magnitude differences as follows:

$$\kappa_{jl}^{shape}(f_j^d, f_l^d) = \exp\left(-\frac{1 - \langle \hat{f}_j^d, \hat{f}_l^d \rangle}{\sigma^{shape}}\right) \tag{7}$$

$$\kappa_{jl}^{mag}(f_j^d, f_l^d) = \exp\left(-\frac{\text{abs}(\|f_j^d\| - \|f_l^d\|)/\|f_j^d\|}{\sigma^{mag}}\right) \tag{8}$$

where $\langle .,. \rangle$ is the dot product. The final similarity value is the product of both kernels:

$$\kappa_{jl}^{NMF}(f_j^d, f_l^d) = \kappa_{jl}^{shape}(f_j^d, f_l^d)\kappa_{jl}^{mag}(f_j^d, f_l^d) \tag{9}$$

With this definition, relative differences based on voxels kinetics are considered. For example, for $d = 1$, all voxels share the same kinetic profile, i.e. $\kappa_{jl}^{shape} = 1$, and differences are only due to relative differences in activity magnitude, captured in $\kappa_{jl}^{mag}$. For $d > 1$, kinetics profile shape can differ between voxels which is additionally captured in $\kappa_{jl}^{shape}$. In all cases we used $\sigma^{shape} = 0.05$ and $\sigma^{mag} = 1$, considering only values for which $\langle \hat{f}_j^d, \hat{f}_l^d \rangle > 0.9$, searching similarities in a 11×11×11 neighborhood and keeping 100 closest voxels based on relative magnitude difference ($k\text{NN}'$):

$$K_{jl}^{NMF} = \begin{cases} \kappa_{jl}^{NMF}, f_l^d \in \text{cluster of } f_j^d \wedge f_l^d \in k\text{NN}' \text{ of } f_j^d \\ 0, \quad \text{otherwise} \end{cases} \tag{10}$$

Finally we also normalize the rows of $K^{NMF}$:

$$\bar{K}^{NMF} = \text{diag}^{-1}[K^{NMF}1_J]K^{NMF} \tag{11}$$

We illustrate the difference between the different kernel definitions in a mouse [$^{18}$F]Fallypride scan of 2 hours, shown in Fig. 1. Kernel values were calculated with: (i) 6 composite frames of 20 min used as voxels features with Euclidian distance differences in the RBFK ($\bar{K}^{comp}$), (ii) using the NMF features $f_j^d$ with the original kernel definition, i.e. normalizing $f_j^d$ by the standard deviation of feature values along the corresponding dimension and using Euclidian distance in the RBFK, or (iii) using the NMF features $f_j^d$ with the proposed kernel definition based on shape and magnitude differences ($\bar{K}^{NMF}$). Kernel values for a voxel in striatum and cerebellum (high contrast regions), and for liver (low contrast) are shown. Range of kernel values in the cerebellum voxel are constrained in the upper range ($> 0.7$) using the Euclidian distance in the RBFK, for both composite frames and NMF features. Using the proposed kernel definition, values are closer to the lower range, ultimately improving contrast. For the striatum, using composite frames fails to find similarity with other voxels. Using NMF features allows to find similarities with other voxels in the striatum, and the proposed kernel increases the number of voxels with similarities. In the liver, a large and relatively homogenous region, kernel values for all cases are in a similar range.

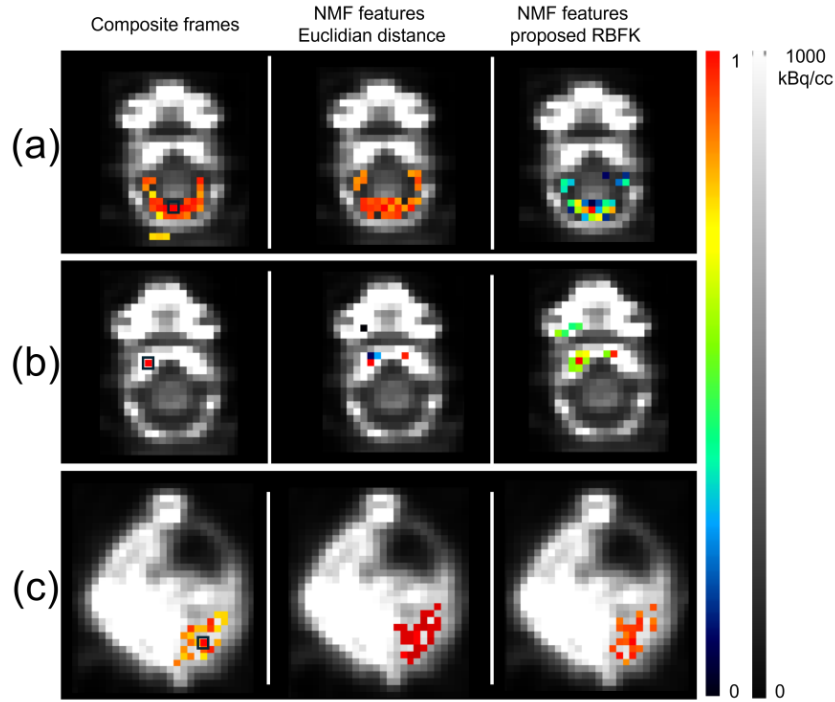

Fig. 1. Example of kernel values using real data from a mouse [$^{18}$F]Fallypride whole body dynamic scan of 2 hours. Kernel values were calculated with composites frames as voxel features and Euclidian distances in the RBFK with $\sigma = 1$ (first column), using the proposed NMF features with Euclidian distance differences in the RBFK with $\sigma = 1$ (middle column), and with NMF features using the proposed kernel definition based on shape and relative magnitude differences with $\sigma^{shape} = 0.05$ and $\sigma^{mag} = 1$ (third column). Kernel values for a voxel (enclosed in black square in first column) in (a) cerebellum (horizontal plane), (b) striatum (horizontal plane), and (c) liver (transverse plane) are shown.

## 2.6 Refinement of kernel matrix coefficients

To further improve contrast and adapt weights in a voxel-by-voxel case, we run a constrained linear least squares optimization for every row of $\overline{K}^{NMF}$. Using voxels features $f_j^d$ as reference data points, we create an underdetermined system of equations where kernel matrix row values are the initial solution:

$$f_j^d \cong F_j \overline{K}_{j*}^{NMF^T} \tag{12}$$

where $F_j$ is created by concatenating in columns the features $f_l^d$ of the $k\text{NN}'$ of $f_j^d$, and $\overline{K}_{j*}^{NMF}$ is the $j$-th row of matrix $\overline{K}^{NMF}$. Then we seek the optimization of the following objective function:

$$\underset{\beta \geq 0}{\operatorname{argmin}} \left\| f_j^d - F_j \beta \right\|^2 \tag{13}$$

where $\beta$ are the refined kernel matrix coefficients for the respective row. We constrain solutions to avoid large deviations from the initial kernel values. To keep track of deviation from the initial kernel values and set a limit on the maximum change allowed, we iteratively find the solution to (13) element by element:

$$\underset{\beta_l^n > 0}{\operatorname{argmin}} g(\beta_l^{n-1}) =$$
$$\underset{\beta_l^n \geq 0}{\operatorname{argmin}} \left\| f_j^d - \left[ (1 - \beta_l^n) R_j^{n-1} + \beta_l^n f_l^d \right] \right\|^2 \tag{14}$$

$$R_j^{n-1} = F_j \beta^{n-1} - \beta_l^{n-1} f_l^d$$

where (14) is designed to maintain the sum to 1 of kernel row elements. Optimization of (14) has a closed form solution, but we clip solutions at every iteration to $\max(\beta_l^n, \beta_l^{n-1} - a\beta_l^{n-1})$ and $\min(\beta_l^n, \beta_l^{n-1} + a\beta_l^{n-1})$, where $0 < a < 1$. We additionally check at every iteration the difference between $\beta_l^n$ and $\overline{K}_{jl}^{NMF}$, and skip to the next element if the difference is above a certain threshold $b$. Algorithm 1 summarizes the kernel matrix coefficients refinement. We used $a = 0.1$, $b = 0.9$, $c = 0.005$, and $maxite = 1000$ in all cases.

**Algorithm 1.** Refinement of kernel matrix coefficients

**Inputs:** Kernel matrix $\bar{K}^{NMF}$, voxels features $f^d$, maximum change per iteration threshold $a$, maximum change from original value threshold $b$, maximum number of iterations $maxite$, threshold in objective function relative change $c$ (stopping criteria)

**Output:** Corrected kernel matrix $\bar{\bar{K}}^{NMF}$

**For** every row in $\bar{K}^{NMF}$
  Set $\beta^{n-1} = \bar{K}^{NMF}_{j*}$, $ite = 0$, $gdiff = inf$
  **If** $ite \leq maxite$ and $gdiff > c$ **then**
    $g^{n-1} = \|f^d_j - F_j \beta^{n-1}\|$
    **For** every element $l$ **in** $\beta$
      $R^{n-1}_j = F_j \beta^{n-1} - \beta^{n-1}_l f^d_l$
      $\beta^{n*}_l = \operatorname{argmin} \|f^d_j - [(1 - \beta^n_l) R^{n-1}_j + \beta^n_l f^d_l]\|^2$
      $\beta^{n*}_l = \max(\min(\beta^{n*}_l, \beta^{n-1}_l + a\beta^{n-1}_l), \beta^{n-1}_l - a\beta^{n-1}_l)$
      $\beta^{n*}_l = \max(\beta^{n*}_l, 0)$
      **If** $abs(\beta^{n*}_l - \bar{K}^{NMF}_{jl})/\bar{K}^{NMF}_{jl} < b$ **then**
        $\beta^{n-1} = (1 - \beta^{n*}_l)[\beta^{n-1}/(1 - \beta^{n-1}_l)]$
        $\beta^n_l = \beta^{n*}_l$
      **Else if**
        $\beta^n_l = \beta^{n-1}_l$
      **End if**
    **End for**
    $gdiff = abs(\|f^d_j - F_j \beta^n\| - g^{n-1})/g^{n-1}$
    $ite = ite + 1;$
  **End if**
  $\bar{\bar{K}}^{NMF}_{j*} = \beta^n$
**End for**

## 2.7 Simulations

To create a realistic 2D phantom, we extracted slice (dynamic) images from heart, brain, kidneys, liver and blood pool of a 1 hour mouse [$^{18}$F]SynVesT-1 scan, filtering the voxels TACs with non local means to smooth the temporal profile while preserving edges [25]. All organ slices were distributed in an image with a size of 128 ×128 pixels, with a pixel size of 0.776 × 0.776 mm. Frames of 12×10 s, 3×20 s, 3×30 s, 3×60 s, 3×150 s, 9×300 s were blurred with a Gaussian full-width at half-maximum (FWHM) of 1.5 mm to simulate scanner point spread function, and frames were forward projected. Poisson noise was added to create 10 dynamic data sets realizations with 300 millions counts in total. Fig. 2 shows the maximum intensity of every voxel along the temporal dimension, and the pixels TACs of every organ.

ML-EM independent frame dynamic reconstruction (IFDR), kernel reconstruction with $\bar{K}^{comp}$, and kernel reconstruction with $\bar{\bar{K}}^{NMF}$ were performed with 300 iterations, including image space resolution recovery [26] with a Gaussian FWHM of 1.5 mm. Bias, variance, and mean squared error (MSE) were calculated as:

$$Bias^2 = \sum_{j=1}^{J} (\bar{x}_j - x_j^{true})^2 / \sum_{j=1}^{J} (x_j^{true})^2 \tag{15}$$

$$Var = \frac{1}{N_r} \sum_{i=1}^{N_r} \sum_{j=1}^{J} (x_j^i - \bar{x}_j)^2 / \sum_{j=1}^{J} (x_j^{true})^2 \tag{16}$$

$$MSE = Bias^2 + Var \tag{17}$$

where $\bar{x}_j$ is the mean over all realizations, $x_j^{true}$ is the ground truth pixel value, and $N_r$ is the number of noisy realizations.

Contrast was calculated as the ratio of mean activity in a hot and cold region, defined in heart by delineation of left ventricle (cold region) and myocardium (hot region), and in kidney by delineation of the inner region of the kidney (cold region) and kidney cortex (hot region). Contrast was calculated in the frame of highest activity for each organ, and plotted with respect to blood pool standard deviation.

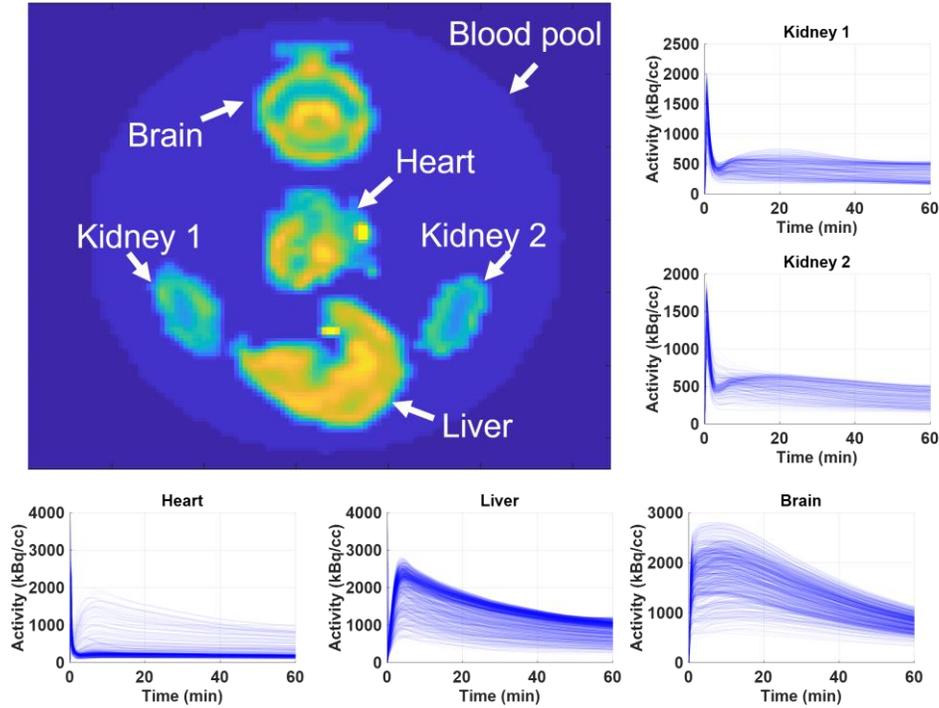

Fig. 2. Simulation phantom created from real mouse [$^{18}$F]SynVesT-1 data, showing all pixels TACs.

2.8 Real data whole body scans

We performed whole body reconstructions of mouse [$^{18}$F]SynVesT-1 scans with high temporal resolution, low dose [$^{11}$C]raclopride, and [$^{18}$F]Fallypride scans. Mouse in [$^{18}$F]SynVesT-1 dynamic scan had an injected dose of 13.6 MBq, reconstructing the first 3 minutes of data in 90 (frames)× 2 (seconds). Mouse in [$^{11}$C]raclopride scan had an injected dose of 1.6 MBq, scan duration was 90 min, with reconstruction frames of 12×10 s, 3×20 s, 3×30 s, 3×60 s, 3×150 s, 15×300 s. Mouse in [$^{18}$F]Fallypride scan had an injected dose of 15.3 MBq, scan duration was 120 min, with reconstruction frames of 12×10 s, 3×20 s, 3×30 s, 3×60 s, 3×150 s, 21×300 s. IFDR, kernel reconstruction using $\bar{K}^{comp}$, and kernel reconstruction using $\bar{\bar{K}}^{NMF}$, were performed with 16 subsets and 32 iterations, with spatially variant resolution modelling [27], and CT based attenuation correction in all cases. All scans were performed in a Siemens Inveon microPET scanner (Siemens Medical Solutions, Inc., Knoxville, USA), reconstructing in a grid of 128×128×159 voxels with a size of 0.776×0.776×0.796 mm in the $x$, $y$, and $z$ directions respectively. The experiments followed the European Ethics Committee recommendations (Decree 2010/63/CEE). Experiments were approved by the Animal Experimental Ethical Committee of the University of Antwerp, Antwerp, Belgium (ECD 2020-71 and 2023-75).

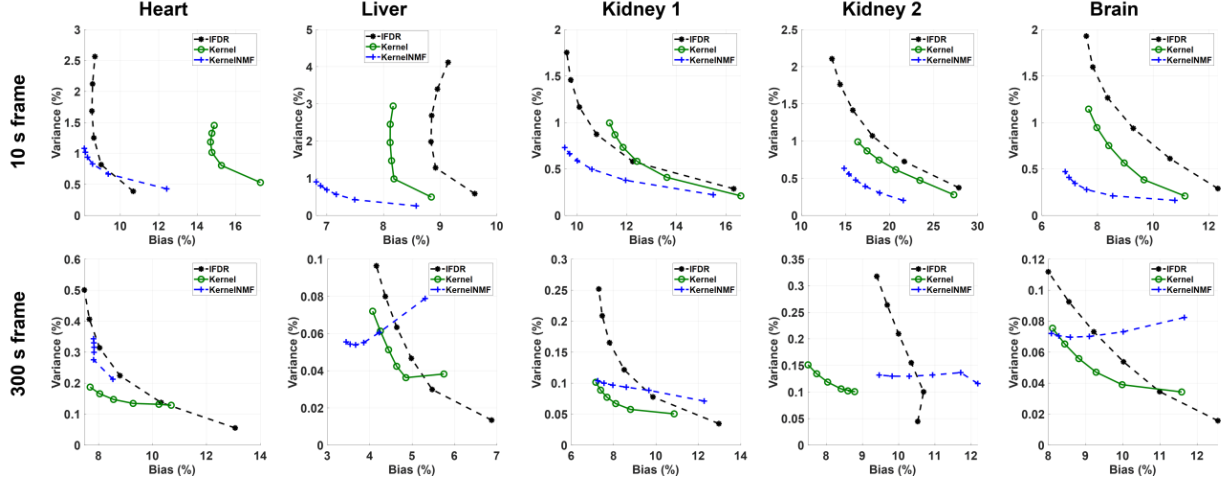

Fig. 3. Bias vs variance plots for the different organs, with IFDR, kernel with $\bar{K}^{comp}$, and kernel with $\bar{\bar{K}}^{NMF}$, from 50 to 300 iterations in 50 iterations steps.

For [$^{18}$F]SynVesT-1, we performed brain kinetic modeling using the heart left ventricle activity as image derived input function (IDIF) [28]. The 1 tissue compartmental model (1TCM) was fitted to voxel TACs to obtain brain parametric maps of $K_1$, as well as parametric maps of fit relative standard error. The whole brain total volume of distribution ($V_T$) was also calculated using the average whole brain TAC. For [$^{11}$C]raclopride, and [$^{18}$F]Fallypride, the simplified reference tissue model (SRTM) 2 [29] was used to generate brain non-displaceable binding potential ($BP_{ND}$) parametric maps, as well as $BP_{ND}$ standard error maps, with cerebellum as reference region. Brain images were initially coregistered to a brain template [30]. For [$^{11}$C]raclopride, the template cerebellum volume of interest (VOI) was used to extract the TAC, while for [$^{18}$F]Fallypride a smaller cerebellum VOI was manually defined to avoid spill over from skull uptake and extrastriatal binding. For all reconstructions the same VOIs were used to extract the IDIF and cerebellum TACs.

## 3. Results

### 3.1 Simulations

Fig. 3 shows the bias vs variance plot for the different organs with IFDR, kernel $\bar{K}^{comp}$ and $\bar{\bar{K}}^{NMF}$. For the early scan 10 s frame, in all cases the kernel methods outperform IFDR, while $\bar{\bar{K}}^{NMF}$ produce the lowest bias and variance. Particularly for the heart the kernel $\bar{K}^{comp}$ has large bias, due to the composite frames failing to capture the heart uptake in early frames. For the late scan 300 s frame differences are smaller. Heart and kidneys, having low activity in these frames, produce better variance with $\bar{K}^{comp}$, while brain and liver have better bias and variance using $\bar{\bar{K}}^{NMF}$.

The MSE of all frames for heart and brain is shown in Fig. 4a. From all frames below 30 seconds (frames 1-18) kernel $\bar{\bar{K}}^{NMF}$ produce the lowest MSE for both organs, while in later, long duration frames the performance is similar between methods.

Contrast vs background STD is shown in Fig. 4b for heart and kidney. IFDR results in the highest contrast but with highest background noise. Kernel $\bar{\bar{K}}^{NMF}$ improves contrast with respect to $\bar{K}^{comp}$ while lowering background noise compared with IFDR.

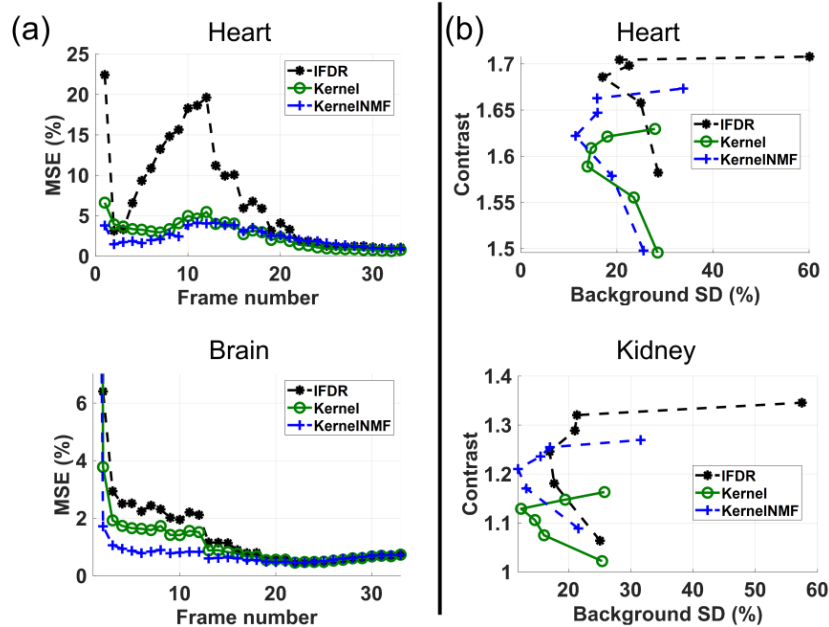

Fig. 4. (a) Mean squared error (MSE) for all frames with IFDR, kernel $\bar{K}^{comp}$, and kernel $\bar{\bar{K}}^{NMF}$, for heart and brain in the simulation phantom. (b) Contrast in heart and liver vs background (blood pool) noise for the different reconstruction methods.

## 3.2 Real data whole body scans

Fig. 5 shows the high temporal resolution reconstruction of different organs in the mouse [$^{18}$F]SynVesT-1 and for [$^{18}$F]Fallypride scans. For [$^{18}$F]SynVesT-1, 2 s frames, noise in IFDR precludes discerning the uptake pattern. Visually, kernel reconstruction with $\bar{K}^{comp}$ improves noise, but oversmooths uptake in regions with fast kinetics such as the heart, vena cava and kidneys. Using $\bar{\bar{K}}^{NMF}$ reduce noise further, while maintaining the high uptake pattern in fast kinetics regions. Similarly for [$^{18}$F]Fallypride, $\bar{\bar{K}}^{NMF}$ outperforms $\bar{K}^{comp}$ in noise reduction and better definition of uptake in fast kinetics regions.

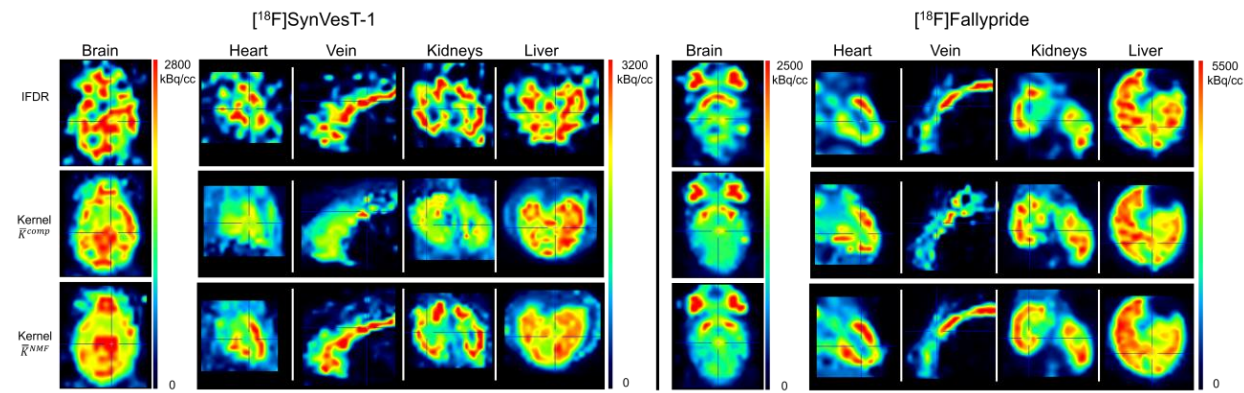

Fig. 5. Example of mouse [$^{18}$F]SynVesT-1 reconstruction of 2 seconds frames of brain (horizontal plane), heart (horizontal), vena cava (sagittal), kidneys (horizontal), and liver (transverse), and for [$^{18}$F]Fallypride brain (horizontal plane, 20 s frame), heart (horizontal, 10s frame), vena cava (sagittal, 10s frame), kidneys (horizontal, 10 s frame), and liver (transverse, 30 s frame), with independent frame dynamic reconstruction (IFDR), and kernel reconstruction using either original kernel matrix ($\bar{K}^{comp}$), or NMF based kernel matrix ($\bar{\bar{K}}^{NMF}$).

Brain kinetic modeling parametric maps of 1TCM $K_1$ for the high temporal reconstruction of [$^{18}$F]SynVesT-1, and [$^{11}$C]raclopride and [$^{18}$F]Fallypride brain $BP_{ND}$ are shown in Fig. 6. For [$^{18}$F]SynVesT-1, noise in brain $K_1$ pattern is reduced using kernel reconstruction with $\bar{K}^{comp}$ compared with IFDR, with further improvement using $\bar{\bar{K}}^{NMF}$. Standard error of the fit is larger than 10% in the entire brain with IFDR, but is reduced to close to 5% using kernel $\bar{K}^{comp}$, and to 2% in the entire brain using $\bar{\bar{K}}^{NMF}$. The whole brain $V_T$ is 10.0, 6.93, and 9.05 ml·cm$^{-3}$ for IFDR, kernel $\bar{K}^{comp}$, and kernel $\bar{\bar{K}}^{NMF}$, respectively.

Similarly for [$^{11}$C]raclopride brain $BP_{ND}$ parametric maps, striatum shape definition is improved using $\bar{\bar{K}}^{NMF}$, compared with IFDR and kernel $\bar{K}^{comp}$, while also greatly reducing $BP_{ND}$ standard error. Mean [$^{11}$C]raclopride striatum $BP_{ND}$ is 3.20, 3.14, and 4.18, and for [$^{18}$F]Fallypride 21.0, 17.1, and 21.2, for IFDR, kernel $\bar{K}^{comp}$, and kernel $\bar{\bar{K}}^{NMF}$, respectively. Striatum $BP_{ND}$ standard error is the lowest using kernel reconstruction with $\bar{\bar{K}}^{NMF}$ in all cases.

## 4. Discussion

Here we propose several modifications to the calculation of the spatial kernel matrix to improve results in whole body PET dynamic reconstruction. Empirically selecting the number of composite frames (i.e. the voxels features) for whole body reconstruction can oversmooth uptake in fast kinetic regions. Increasing the number of composite frames can help to capture faster kinetics information, but at the expense of noise increase, and consequently hindering calculation of voxels features similarities. Additionally, empirically setting a threshold on the RBFK values to improve contrast can compromise noise reduction, especially when noise is present in the voxels similarities calculation. To handle these issues, we propose to optimally calculate the minimum dimension of voxels features necessary to capture all kinetic information using nonnegative matrix factorization. With this new definition of the voxels features, we also propose new voxels similarities calculation based on relative differences, which adapts better to regions with highly different uptake. Finally, we propose a new refinement to the kernel matrix coefficients to improve contrast while maintaining noise reduction. Simulations and real data reconstructions show improvements using our method compared with the original kernel reconstruction.

Instead of using composite frames, here we use the whole IFDR to calculate voxels features. This is key to capture all kinetic information present in the data. Composite frames, although reducing noise, average the kinetic information contained in the respective time duration of the frame. In our tests we found that for many radiotracers, uptake of fast kinetics organs/regions such as the heart, kidneys and veins, is present only for few seconds/minutes in the image, and therefore "disappear" in composite frames of more than a few minutes duration. Thus, voxels similarity are calculated with uniform uptake images in these regions, which create spatial basis function that do not reflect the shape of the organ uptake. This is then translated in oversmoothing and artefacts in these regions when kernel reconstruction is performed. Although in theory the entire IFDR voxel TAC can be used as the voxel feature to capture the fast kinetics, the high noise in the voxel TAC hinders calculation of voxels similarities. Therefore, in order to capture all kinetic information while also maintaining noise reduction, NMF is used in the IFDR to identify the finite number of kinetics profiles present in the data, producing low noise images of the different uptake spatial patterns which can be used as voxels features.

NMF is ideally suited for analysis of PET data due to its non-negativity constraint. Other dimensionality reduction methods, such as principal component analysis [31], singular value decomposition [32], and independent component analysis [33], have been used to analyse or reconstruct PET data, but these methods can produce negative values, violating the nonnegative nature of PET data. NMF on the other hand explicitly optimize for nonnegative results.

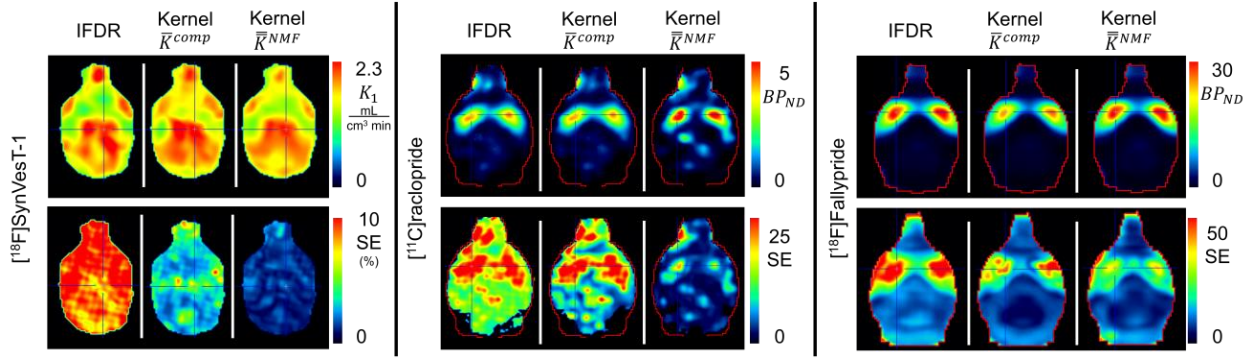

Fig. 6. Horizontal plane slices of 1TCM $K_1$ parametric maps of [$^{18}$F]SynVesT-1 mouse brain and $K_1$ relative standard error (SE), and [$^{11}$C]raclopride and [$^{18}$F]Fallypride brain $BP_{ND}$ parametric maps and $BP_{ND}$ absolute SE. Kinetic modeling performed using independent frame dynamic reconstruction (IFDR), and kernel reconstruction using either original kernel matrix ($\bar{K}^{comp}$), or NMF based kernel matrix ($\bar{\bar{K}}^{NMF}$).

Apart from the minimum volume regularized NMF [17], we also tested the original NMF with multiplicate rule optimization [34], and orthogonal NMF [35], both with nonnegative ICA initialization, but the first method gave the best results in terms of stability among runs, which improved results in the rank selection algorithms. Moreover, reasonable scaling of pure TACs ($W$) naturally resulted using minimum volume regularized NMF with the $H^T 1_r \leq 1$ constraint, while scaling of $W$ with the other NMF methods had to be calculated afterwards, sometime resulting in suboptimal results.

NMF has been previously used for PET reconstruction [36], but the previously proposed method requires training of convolutional neural networks, therefore requiring fine tuning of training parameters. Importantly that method was tested only in single organ reconstruction, and the rank of the NMF decomposition is a user defined parameter, limiting its application to whole-body PET dynamic reconstruction. To our knowledge, our work is the first using NMF in the context of PET kernel dynamic reconstruction for whole body PET, also addressing the problem of automatic rank selection for PET data.

Automatic rank selection allows to optimally select the dimensionality of the voxels features by calculating the minimum number of kinetic profiles that can describe the data while also discarding noise. On the other hand, empirically dividing the data in composite frames can introduce unnecessary noise in the voxels features. For example, in the low dose [$^{11}$C]raclopride scan it was determined that a feature of one dimension was optimal for the brain and blood pool region, since, due to high noise, 2 or more components fitted the noise. Conversely, each of the empirically selected composite frames in the [$^{11}$C]raclopride scan contained high noise, which translated in poor calculation of voxels similarities.

Our method requires segmentation of the PET image in order to better discern the different kinetic profiles present in every body region. Applying NMF directly to the whole body is possible, but including voxels TACs with highly different kinetic profiles in the same dataset can hinder distinction of subtle differences in TACs kinetics. However, is important to note that "perfect" segmentation is not necessary, as long as regions with highly different kinetics are separated. Performing NMF using all body voxels dynamic data usually fails to separate kinetic profiles that are subtly different. For example, including the bladder and brain in the same dataset can separate the profiles of both organs, but the high and low specific binding profiles in the brain might not be differentiated. This is in part due to the "well spread" condition in NMF [15], which states that data points need to be well distributed in the convex hull spanned by the columns of $W$. In our experiments with data from PET dynamic scans using the NMF method in [17], this condition was rather relaxed, and small regions embedded in large clusters could still be differentiated. For example, in the thoracic cluster containing the heart, sometimes part of the liver wall was segmented together due to partial volume effect. Although less than 5% of the voxels in the cluster belonged to the liver, the liver component could still be differentiated. Other clustering methods can be used, such as anatomical based clustering if anatomical imaging is also available, but in general undersegmentation is preferable to oversegmentation in order to create datasets with larger number of voxels TACs. In the case of single organ imaging (e.g. brain), the clustering step can be omitted.

Another key aspect of our method is the use of relative differences to calculate the similarity between voxels. In the original kernel method the Euclidian distance between features is used to calculate differences, which are then scaled by a uniform $\sigma$ in the RBFK [3]. Although features are normalized by their standard deviation, the magnitude of feature values can be highly different within the feature dimension. Therefore, the scale of the features differences will also be different between regions, making the use of a single $\sigma$ value suboptimal. For example, in our [$^{18}$F]Fallypride dataset, several voxel similarities were found in the low activity region cerebellum, but none in the high activity striatum. Increasing the value of $\sigma$ allows to find more voxel similarities in striatum, but at the expense of including outliers in the cerebellum similarities. By using relative differences, a single $\sigma$ value in the RBFK produce equally well results in regions with different features magnitude.

Relative differences are possible to calculate using the NMF features interpreted as vectors describing shape and magnitude of the kinetic profiles. Euclidian distances can also be used to calculate differences between NMF features, but then the same issue of $\sigma$ scaling would arise.

As with any regularization method, using the kernel method can oversmooth the activity in small high contrast regions. In the kernel method this is due to having kernel matrix coefficients with high weight for voxels that have highly different activity magnitude. Originally this was corrected by setting a threshold in the RBFK values to remove weights for voxels with large features differences. This can result in kernel matrix rows with very few, or even a single weight, therefore compromising noise reduction. The algorithm we propose here to correct for this utilize the same voxels features used to find voxels similarities as ground truth reference to adjust the kernel matrix coefficients. To avoid setting coefficients to zero, the algorithm bounds the solution to have a maximum difference with respect to the original kernel matrix coefficients. This helped to improve contrast while maintaining noise reduction in the kernel reconstruction. The proposed algorithm iteratively adjusts individual matrix coefficients to keep track of their change and skip results when the change is too large. We also tested solving (13) using projected gradient descent, with projection onto the nonnegative orthant and onto the $\sum_l \beta_l = 1$ hyperplane, but in some cases many coefficients were set to zero with this method.

Simulations with ground-truth reference activity show improvement in bias-variance trade-off, and in contrast, using $\overline{\overline{K}}^{NMF}$ over $\overline{K}^{comp}$, specially in early, short duration frames, for organs with different kinetic profiles. In the segmented heart image used for the simulation, some liver spill-over is observed in later frames. Using NMF on IFDR data shows separation of the heart and spill-over regions, while composite frames fail to capture this separation. This translates in artifacts in the heart image, shown in the large bias in early frames. In later frames when kinetics are slower, composite frames are a good representation of the true noise-less activity, resulting in similar performance between methods.

In high temporal reconstruction of real [$^{18}$F]SynVesT-1 data, despite having greater amount of composite frames where the fast kinetics uptake is present, the use of $\overline{K}^{comp}$ oversmooths uptake in these regions. This is particularly detrimental when an image derived input function is used for kinetic modeling, where the heart left ventricle or vena cava activity is usually used [37]. In slower kinetics regions, such as the brain and liver, oversmoothing is less present. Using $\overline{\overline{K}}^{NMF}$ maintains contrast in fast kinetics organs, while further reducing noise. The voxels features calculated using the IFDR with NMF capture fast and slow kinetics equally well, even when few voxels exhibit the fast kinetics profile for short time duration, such as in the mouse heart. Reducing noise for high temporal resolution PET reconstruction can be beneficial for its use with newly developed kinetic models which require this high resolution [38].

The proposed method also improve reconstruction for low dose PET dynamic imaging, as demonstrated in the [$^{11}$C]raclopride scan. This is desirable either to avoid high radiation burden to the patient, or to get closer to tracer concentration doses in preclinical scans [39]. Kernel reconstruction with composite frames only modestly reduced noise in this case due to the use of composite frames with high noise. NMF voxels features on the other hand are optimally calculated to discard the noise component, which results in lower noise kernel reconstructions.

In the [$^{18}$F]Fallypride real data, fast kinetics regions presented artifacts using kernel reconstruction with composite frames. This was corrected using the proposed method, with clear definition of heart, kidneys and veins uptake.

Using the proposed method, brain kinetic modeling of [$^{18}$F]SynVesT-1 data with the 1TCM, and of [$^{11}$C]raclopride and [$^{18}$F]Fallypride with the SRTM, produced parametric images with better brain structures definition, and parameters with lower standard error compared with the original kernel method. For [$^{18}$F]SynVesT-1 using the original kernel method, oversmoothing of the IDIF results in lower brain $V_T$ compared with both IFDR and the NMF kernel reconstruction. This effect is also observed in [$^{18}$F]Fallypride striatum $BP_{ND}$, where lower values are observed using the original kernel method compares with IFDR and NMF kernel due to oversmoothing of the cerebellum TAC. The NMF kernel reconstruction maintains high contrast in parametric images while also reducing noise.

As with the original kernel matrix, the (functional) NMF spatial kernel matrix can be used together with an anatomical kernel matrix (e.g. calculated from MRI) [9], and/or with a temporal kernel matrix [12] to further reduce noise.

Although we used small animal reconstructions to show the improvements of our method, clinical PET will also benefit from the proposed method, particularly with recent introduction of whole body PET scanners [40]. High contrast at the voxel level is also desirable, for example, in new high resolution brain PET scanners [41], where noise reduction methods can smooth differences in small regions. Reconstructions in "traditional" clinical PET scanners can also be brought closer to low noise reconstructions form high sensitivity scanners using the proposed method here.

## 5. Conclusions

In this work we designed a new functional spatial kernel matrix for whole body dynamic PET reconstruction, using nonnegative matrix factorization to calculate the voxels features. The proposed method improves noise reduction, while maintaining high contrast, compared with the original kernel matrix. This is observed in body regions of both fast and slow radiotracer kinetics. Simulations and real data sets reconstructions show the improvements of our method, particularly in small regions of high contrast and fast kinetics where the original kernel method produce suboptimal results. Kinetic modelling parametric maps also show reduced noise and parameters with lower standard error using the proposed kernel matrix. Whole body dynamic PET with NMF kernel reconstruction can help to explore whole body kinetic modelling with reduced bias, possible now with state-of-the-art PET scanners.